\documentclass[conference]{IEEEtran}

\usepackage[T1]{fontenc}
\usepackage{graphicx}
\usepackage{cite}
\usepackage{caption}
\usepackage{subcaption}
\usepackage{epstopdf}
\usepackage{lipsum}

\usepackage{overpic}
\usepackage{amssymb}
\usepackage{amsmath}
\usepackage{amsthm}
\usepackage{array}
\usepackage{color}
\usepackage{url}
\usepackage{hyperref}

\IEEEoverridecommandlockouts

\theoremstyle{plain}
\newtheorem{theorem}{Theorem}

\newcommand{\vect}[1]{\mathbf{#1}}

\def\Real{\mathbb{R}}

\def\kron{\otimes}

\def\Htran{\mbox{\tiny $\mathrm{H}$}}
\def\Ttran{\mbox{\tiny $\mathrm{T}$}}
\def\CN{\mathcal{N}_{\mathbb{C}}} 
\def\imagunit{\mathsf{j}} 

\def\krx{{\bf k}}
\def\ktx{\boldsymbol{\kappa}}

\begin{document}

\title{Holographic MIMO Communications Under Spatially-Stationary Scattering \vspace{-0.4cm}}

\author{
\IEEEauthorblockN{Andrea Pizzo\IEEEauthorrefmark{1}\IEEEauthorrefmark{2}, Thomas Marzetta\IEEEauthorrefmark{1,2}
Luca Sanguinetti\IEEEauthorrefmark{2}
\thanks{L. Sanguinetti was partially supported by the Italian Ministry of Education and Research (MIUR) in the framework of the CrossLab project.}}
\IEEEauthorblockA{\IEEEauthorrefmark{1}\small{Department of Electrical and Computer Engineering, New York University, USA (\{andrea.pizzo,tom.marzetta\}@nyu.edu)}}
\IEEEauthorblockA{\IEEEauthorrefmark{2}\small{Dipartimento di Ingegneria dell'Informazione, University of Pisa, Italy (luca.sanguinetti@unipi.it)}\vspace{-0.6cm}}
}

\maketitle

\begin{abstract}
Holographic MIMO is a spatially-constrained MIMO system with a massive number of antennas, possibly thought of, in its ultimate form, as a spatially-continuous electromagnetic aperture. Accurate and tractable channel modeling is critical to understanding the full potential of this technology. This paper considers arbitrary spatially-stationary scattering and provides a 4D plane-wave representation in Cartesian coordinates, which captures the essence of electromagnetic propagation and allows to evaluate the capacity of Holographic MIMO systems with rectangular volumetric arrays. The developed framework generalizes the virtual channel representation, which was originally developed for uniform linear arrays. 
\vspace{-0.1cm}
\end{abstract}


\IEEEpeerreviewmaketitle

\section{Introduction}
Holographic MIMO refers to arrays with a massive number of antennas in a compact space~\cite{BJORNSON20193}. In its asymptotic form, it can be thought of as a spatially-continuous electromagnetic aperture that actively generates beamformed radio signals. Maximal exploitation of this technology necessitates accurate yet tractable modeling of the MIMO channel matrix ~\cite{Marzetta2018}. MIMO channel models are either deterministic or stochastic.  Deterministic models are accurate, but valid only for the considered propagation conditions. Unlike deterministic models, stochastic approaches are independent of a particular propagation environment. A widely used model is the i.i.d. Rayleigh fading channel model, which assumes that the channel realizations observed at the antennas are statistically independent. This model is widely used because of its analytical tractability. However, it is physically meaningful only if a half-wavelength spaced one-dimensional (1D) array is deployed in isotropic scattering~\cite{JSAC}. With two-dimensional (2D) and three-dimensional (3D) arrays, the closest physically-tenable model for i.i.d. Rayleigh fading is the Clarke's model~\cite{Marzetta2018,JSAC}, which correctly exhibits spatial correlation even with isotropic scattering. Alternatively, there are the parametric physical models inspired by array signal processing, which tend to leave out underlying characteristics of electromagnetic propagation. 

In this paper, we consider arbitrary spatially-stationary scattering and provide a Fourier plane-wave representation in Cartesian coordinates \cite{MarzettaNokia}, which captures the essence of electromagnetic propagation and allows to evaluate the capacity of Holographic MIMO systems with rectangular volumetric arrays. The developed model generalizes the virtual channel representation pioneered in~\cite{Sayeed2002} for 1D arrays. When sampled in space, our model constitutes a \emph{physics-based low-rank approximation} of the channel, which belongs to the unitary-independent-unitary (UIU) models~\cite{Tulino2005}.

\section{Continuous-Space Electromagnetic Channel for LoS propagation}

Consider a wireless communication between two continuous-space rectangular volumes and assume that it takes place through scalar waves (i.e., no polarization) propagating into a 3D homogeneous, {isotropic}, and infinite medium characterized by a permeability $\mu$ and permittivity $\epsilon$.  
In any frequency-flat channel, the electric field $e(\vect{r})$ measured at the receiver is related to the electric-current volume density $j(\vect{s})$  at the source as
\begin{equation} \label{conv_model}
e(\vect{r})  =  \int_{\mathcal{V}_s} e(\vect{r},\vect{s}) \, d\vect{s} =  \int_{\mathcal{V}_s} h(\vect{r},\vect{s}) j(\vect{s}) \, d\vect{s}, \quad {\vect{r} \in \mathcal{V}_r}
\end{equation}
where the field $h(\vect{r},\vect{s})$ represents the \emph{spatial impulse response} of the channel at $\vect{r}$ to an impulse (point-source) of electric current applied at $\vect{s}$.
Here, $e(\vect{r},\vect{s})$ is the electric field generated by a point source of electric-current at point $\vect{s}$ and measured at an arbitrary point $\vect{r}$. Hence, the wireless channel can generally be viewed as a linear and space-variant system with constitutive relation given by \eqref{conv_model}. Next, we start considering a line-of-sight (LoS) scenario, i.e., free-space propagation.

\begin{figure*}[t!]
\vspace{-0.5cm}
\begin{align}\tag{11} \label{Fourier_planewave}
h(\vect{r},\vect{s})  =  \frac{1}{(2\pi)^2}
\iiiint_{-\infty}^{\infty} a_r(\krx,\vect{r})
H_a(k_x,k_y,\kappa_x,\kappa_y)  a_s(\ktx,\vect{s}) \, dk_xdk_y d\kappa_xd\kappa_y
\end{align} \vspace{-0.2cm}
\hrule\vspace{-0.5cm}
\end{figure*}

\subsection{Green function}

In LoS scenario, the electric-field $e(\vect{r},\vect{s})$ is given by \cite{ChewBook}
\begin{equation}\label{electric-field}
e(\vect{r},\vect{s}) = -\imagunit \kappa \eta g(\vect{r},\vect{s}) j(\vect{s}) 
\end{equation}
where ${\eta = \sqrt{\mu/\epsilon}}$ is the intrinsic impedance and 
\begin{equation}  \label{Green}
g(\vect{r},\vect{s}) =    \frac{e^{\imagunit \kappa \|\vect{r}-\vect{s}\|}}{4\pi\|\vect{r}-\vect{s}\|}
\end{equation}
is the scalar Green function that represents the solution
to the inhomogeneous scalar Helmholtz equation $(\nabla^2  + \kappa^2) g(\vect{r},\vect{s}) = -\delta(\vect{r}-\vect{s})$ \cite[Eq.~1.3.34]{ChewBook}.
Notice that $\kappa=\omega/c$ is the wavenumber with  $\omega$ being the radiation frequency  and $c = 1/\sqrt{\epsilon \mu}$  the speed of light. 
From \eqref{conv_model}, the channel impulse response under LoS propagation conditions is $h(\vect{r},\vect{s}) = -\imagunit \kappa \eta g(\vect{r},\vect{s})$.

The Green function in \eqref{Green} describes a scalar \emph{spherical-wave} that propagates radially from point $\vect{s}$ in every directions. 
Next, we show that a spherical wave can be decomposed as a superposition of plane-waves.
This is instrumental to derive an \emph{exact} plane-wave representation of $h(\vect{r},\vect{s})$ in LoS conditions. This representation will be extended in Section~\ref{SectionIII} to model $h(\vect{r},\vect{s})$ under spatially-stationary scattering conditions.  

\subsection{Fourier plane-wave representation}
%
%
%
By using the Weyl's identity \cite[Eq.~2.2.27]{ChewBook}, we have that
\begin{equation} \label{Weyl_identity}
\frac{e^{\imagunit \kappa R}}{R}= \frac{\imagunit}{2\pi}  \iint_{-\infty}^{\infty} \frac{e^{\imagunit \big( \kappa_x x + \kappa_y y + \kappa_z(\kappa_x,\kappa_y) z\big)}}{\kappa_z(\kappa_x,\kappa_y)}  \, d\kappa_xd\kappa_y, \, z>0
\end{equation}
where  $R = \sqrt{x^2 + y^2 + z^2}$ and
\begin{equation} \label{kappaz_tx}
\kappa_z(\kappa_x,\kappa_y) = 
\begin{cases}
\gamma(\kappa_x,\kappa_y), &  (\kappa_x,\kappa_y) \in \mathcal{D}  \\
\imagunit |\gamma(\kappa_x,\kappa_y)|, & (\kappa_x,\kappa_y) \in \mathcal{D}^c
\end{cases}
\end{equation}
where $\gamma(\kappa_x,\kappa_y)$ is defined as
\begin{equation} \label{gamma}
\gamma(\kappa_x,\kappa_y) =  \sqrt{ \kappa^2 - \kappa_x^2 - \kappa_y^2} 
\end{equation}
and it is real-valued within the support
\begin{equation} \label{disk}
\mathcal{D} = \{ (\kappa_x,\kappa_y)\in\Real^2 : \kappa_x^2 + \kappa_y^2 \le \kappa^2\}
\end{equation}
and imaginary-valued on the complementary set $\mathcal{D}^c$ of $\mathcal{D}$. In \eqref{kappaz_tx}, $\Re \{\kappa_z\}>0$ and $\Im\{\kappa_z\}>0$ for $(\kappa_x,\kappa_y)\in\Real^2$ in order to ensure the radiation condition at infinity \cite[Ch.~2]{ChewBook}.
By using the identity~\eqref{Weyl_identity} into \eqref{Green}, the electric field $e(\vect{r},\vect{s})$ is given by~\eqref{electric-field} with channel impulse response in the form of
\begin{equation}\label{small-scale-LoS}
h(\vect{r},\vect{s}) =   \frac{\kappa \eta}{2(2\pi)^2}  \iint_{-\infty}^{\infty}  \frac{e^{\imagunit  \ktx ^{\Ttran} (\vect{r} - \vect{s})}}{\kappa_z(\kappa_x,\kappa_y)}  \, d\kappa_xd\kappa_y, \quad \! r_z>s_z\!
\end{equation}
where
\begin{equation}
 \ktx = [\kappa_x,\kappa_y,\kappa_z(\kappa_x,\kappa_y)]^{\Ttran}
\end{equation}
is the wave vector which corresponds to the source propagation direction $\hat \ktx = \ktx/||\ktx||$ of the radiated field.
From~\eqref{small-scale-LoS}, it thus follows that the channel response can be represented  \emph{exactly} as a superposition of \emph{plane-waves} $e^{\imagunit  \ktx ^{\Ttran} (\vect{r} - \vect{s})}$ 
that are parameterized by the horizontal wavenumbers $(\kappa_x,\kappa_y)$. This result is valid irrespective of the distance between the arbitrary points $\bf{r}$ and $\bf{s}$. Therefore, it is true also in the  near-field.\footnote{Elementary treatments of electromagnetic wave propagation teach that the electric field can be represented with a plane-wave in the far-field, i.e., when $\bf{r}$ is at a sufficient number of wavelenghts from $\bf{s}$. It is less obvious that a plane-wave representation is also possible in the near-field \cite{MarzettaNokia}.}

An inspection of \eqref{small-scale-LoS} reveals that an infinite number of plane-waves are radiated by the point source at $\vect{s}$. Since no material objects are present in LoS conditions, there will be an infinite number of plane-waves at the receiver whose propagation directions $\hat \krx = \krx/||\krx||$ with
\begin{equation}
 \krx = [k_x,k_y,k_z(k_x,k_y)]^{\Ttran}
\end{equation}
 are in one-to-one correspondence to the ones at the source, i.e., $\hat \krx = \hat \ktx$.
Under LoS propagation conditions, \eqref{small-scale-LoS} can thus be written as the \emph{4D plane-wave representation} in \eqref{Fourier_planewave} where
\setcounter{equation}{11} 
\begin{align}\label{channel-model}
H_a(k_x,k_y,\kappa_x,\kappa_y) = \frac{\kappa \eta}{2} \frac{\delta(k_y-\kappa_y)\delta(k_x-\kappa_x)}{\kappa_z(\kappa_x,\kappa_y)}\end{align}
and
\begin{align}\label{plane-wave-tx}
a_s(\ktx,\vect{s}) & = e^{-\imagunit  \ktx ^{\Ttran} \vect{s}} = e^{-\imagunit \big( \kappa_x s_x + \kappa_y s_y + \kappa_z(\kappa_x,\kappa_y) s_z\big)} \\ \label{plane-wave-rx}
a_r(\krx,\vect{r}) & = e^{\imagunit  \krx ^{\Ttran} \vect{r}} = e^{\imagunit \big( k_x r_x + k_y r_y + \kappa_z(k_x,k_y) r_z\big)}
\end{align}
are the source and receive plane-waves, respectively. Notice that delta functions in~\eqref{channel-model} ensure the LoS condition $\hat \krx = \hat \ktx$. 

The plane-wave representation in \eqref{Fourier_planewave} suggests that $h(\vect{r},\vect{s})$ can be decomposed in three terms. 
In particular, $a_s(\ktx,\vect{s})$ is the source response that maps the excitation current at any point $\vect{s}$ to the source propagation direction $\hat \ktx$ of the radiated field. Similarly, $a_r(\krx,\vect{r}) $ is the receive  response that maps the receive propagation direction $\hat \krx$ of the receive field to the induced current  at $\vect{r}$. The complex coefficient $H_a(k_x,k_y,\kappa_x,\kappa_y)$ is the angular response that maps the source direction $\hat\ktx$ into the  receive direction $\hat\krx$. Notice that $H_a(k_x,k_y,\kappa_x,\kappa_y)$ acts only on the horizontal coordinates since the longitudinal coordinates $(k_z,\kappa_z)$ follow directly.

Notice from \eqref{plane-wave-tx} and \eqref{plane-wave-rx} that the plane-waves coincide with phase-shifted versions of 2D spatial-frequency Fourier harmonics $e^{-\imagunit ( \kappa_x s_x + \kappa_y s_y )}$ and $e^{\imagunit ( k_x r_x + k_y r_y)}$, respectively. 
In light of this, we call \eqref{Fourier_planewave} the \emph{4D Fourier plane-wave representation} of $h(\vect{r},\vect{s})$.

\section{Continuous-Space Electromagnetic Channel for Spatially-Stationary Scattering}\label{SectionIII}
The above analysis focuses on LoS scenarios. We now account for the presence of a scatterer with arbitrary shape between the two rectangular volumes $\mathcal{V}_s$ and $\mathcal{V}_r$.
%
Particularly, we assume that it can be modelled as a circularly-symmetric, complex-Gaussian, stationary spatial random field. Under these conditions, we derive an integral plane-wave representation of $h(\vect{r},\vect{s})$, which is valid for infinite rectangular volumes $\mathcal{V}_r$ and $\mathcal{V}_s$ of finite thickness, i.e., ${(r_x,r_y)\in \mathbb{R}^2}$ and ${(s_x,s_y) \in \mathbb{R}^2}$, respectively.




\subsection{Fourier plane-wave spectral representation}



Similar to stationary time-domain random processes~\cite{VanTreesBook}, Theorem 1 below provides the Fourier spectral representation of a stationary spatial random field $h(\vect{r},\vect{s})$ of electromagnetic nature. 
Details can be found in the extended version.
\begin{theorem} \label{th:spectral}
The spatial impulse response $h(\vect{r},\vect{s})$ of a stationary channel over volumetric arrays with ${(r_x,r_y)\in \mathbb{R}^2}$ and ${(s_x,s_y) \in \mathbb{R}^2}$ is exactly given by the 4D Fourier plane-wave representation in \eqref{Fourier_planewave} with angular response
\begin{align} 
 \label{scattering_response_nlos}
\hspace{-0.4cm}H_a(k_x,k_y,\kappa_x,\kappa_y) \!=\!  \sqrt{S(k_x,k_y,\kappa_x,\kappa_y)}  W(k_x,k_y,\kappa_x,\kappa_y) \!\!
\end{align}
where 
\begin{equation} \label{psd}
\!\!S(k_x,k_y,\kappa_x,\kappa_y)   = \Big(\frac{\kappa \eta}{2}\Big)^2 \frac{A^2(k_x,k_y,\kappa_x,\kappa_y) }{\gamma(k_x,k_y) \gamma(\kappa_x,\kappa_y)}
\end{equation}
with $(k_x,k_y,\kappa_x,\kappa_y)\in\mathcal{D}\times \mathcal{D}$ is the angular power spectral density of $h(\vect{r},\vect{s})$ with $A(\cdot)$ being a non-negative function and $W(\cdot)$ is a complex white-noise  field with unit variance. Also, we have that $\kappa_z(\kappa_x,\kappa_y) = \gamma(\kappa_x,\kappa_y)$ and $k_z(k_x,k_y) = \gamma(k_x,k_y)$ in \eqref{Fourier_planewave}.
\end{theorem}
We notice that $h(\vect{r},\vect{s})$ in Theorem 1 is fully described by the angular power spectral density $S(k_x,k_y,\kappa_x,\kappa_y)$. Building on \cite{Marzetta2018,JSAC}, it can be shown that, for spatially-stationary electromagnetic channels, $S(k_x,k_y,\kappa_x,\kappa_y)$ must be circularly band-limited with $(k_x,k_y,\kappa_x,\kappa_y)\in\mathcal{D}\times\mathcal{D}$. As seen in~\eqref{kappaz_tx}, plane-waves associated with propagation directions outside the support $\mathcal{D}\times\mathcal{D}$ have an imaginary-valued wavenumber along the longitudinal coordinates $\kappa_z$ and $k_z$. Hence, their power decays exponentially fast in space. These plane-waves are called evanescent  and  lead to a non-stationary $h(\vect{r},\vect{s})$.

We finally notice that $S(k_x,k_y,\kappa_x,\kappa_y)$ in \eqref{psd} is uniquely described by the so-called spectral factor $A(k_x,k_y,\kappa_x,\kappa_y)$ that physically accounts for the angular selectivity of the scattering, i.e., the power transfer between source and receive propagation directions. The factors $\gamma(\cdot)$ at the denominator of \eqref{psd} are the Jacobians of the plane-waves parametrizations with respect to the horizontal wavenumber coordinates $(\kappa_x,\kappa_y)$ and $(k_x,k_y)$~\cite{JSAC}.
An isotropic scattering is characterized by a radially-symmetric spectral factor, which is given by $ A(k_x,k_y,\kappa_x,\kappa_y) = \tilde A = \frac{2}{\kappa \eta} (\frac{2 \pi}{\sqrt{\kappa}})^2$ when the overall power of $h(\vect{r},\vect{s})$ is normalized to $1$ \cite{JSAC}.

\begin{figure*}
\vspace{-0.5cm}
\begin{equation}  \tag{24}\label{Fourier_series}
h(\vect{r},\vect{s}) \approx \hspace{-0.35cm}\mathop{\sum}_{(\ell_x,\ell_y)\in\mathcal{E}_r} \mathop{ \sum}_{(m_x,m_y)\in \mathcal{E}_s}  \hspace{-0.35cm}H_a(\ell_x,\ell_y,m_x,m_y) \underbrace{a_{r}(\ell_x,\ell_y,\vect{r})  a_{s}(m_x,m_y,\vect{s})}_{\phi(\ell_x,\ell_y,m_x,m_y,\vect{r},\vect{s})}
\end{equation}\vspace{-0.2cm}
\hrule\vspace{-0.5cm}
\end{figure*}

\subsection{Fourier plane-wave series expansion}\label{Fourier-series}
Theorem 1 provides us with an \emph{exact} closed-form representation of a spatially-stationary channel over infinite $z-$planes, i.e., ${(r_x,r_y)\in \mathbb{R}^2}$ and ${(s_x,s_y) \in \mathbb{R}^2}$. Theorem 2 below provides a series expansion that is a good approximation of the Fourier plane-wave spectral representation over volumes of practical dimensions (compared to the wavelength). 
Details will be provided in the extended version. We limit to observe that Theorem 2 extends \cite[Sec. V]{JSAC}. 

To state the main result of Theorem 2 below, we assume that the two rectangular volumes $\mathcal{V}_s$ and $\mathcal{V}_r$ have finite side lengths $L_{r,x},L_{r,y}, L_{r,z} $ and $L_{s,x} ,L_{s,y},L_{s,z}$, respectively. With a slight abuse of notation, we call 
\begin{align} \label{plane_wave_series}
\!\!\!a_{s}(m_x,m_y,\vect{s}) & = e^{-\imagunit \Big(\frac{2\pi}{L_{s,x}} m_xs_x + \frac{2 \pi}{L_{s,y}} m_ys_y + \gamma_s(m_x,m_y) s_z \Big)}\!
\end{align}
with 
\begin{equation}
\!\!\!\gamma_s(m_x,m_y) = \sqrt{\kappa^2-(2\pi m_x/L_{s,x})^2-(2\pi m_y/L_{s,y})^2}
\end{equation}
 the discretized plane-wave harmonics as obtained from $a_s(\ktx,\vect{s})$ in~\eqref{plane-wave-tx} by evaluating $(\kappa_x,\kappa_y)$ at $({2\pi m_x}/L_{s,x},{2\pi m_y}/L_{s,y})$. Consequently, we have that $\kappa_z = \gamma_s(m_x,m_y)$. Similarly, we call $a_{r}(\ell_x,\ell_y, \vect{r}) $
\begin{align} \label{plane_wave_series}
a_{r}(\ell_x,\ell_y, \vect{r}) = e^{\imagunit \Big(\frac{2\pi}{L_{r,x}} \ell_xr_x + \frac{2 \pi}{L_{r,y}} \ell_yr_y + \gamma_r(\ell_x,\ell_y) r_z \Big)}
\end{align}
the discretization of  $a_r(\krx,\vect{r})$ in~\eqref{plane-wave-rx} with $\gamma_r(\ell_x,\ell_y)$ obtained accordingly. 
Since $h(\vect{r},\vect{s})$ is circularly-bandlimited with $(k_x,k_y,\kappa_x,\kappa_y)\in\mathcal{D}\times\mathcal{D}$, from~\eqref{disk} the plane-wave harmonics are non-zero only within the lattice ellipse (e.g. \cite[Fig. 1]{SPAWC})
\begin{align}\label{epsilon_s}
\hspace{-0.4cm}\mathcal{E}_s &= \{(m_x,m_y)\!\in\!\mathbb{Z}^2 \!:\! (m_x \lambda/L_{s,x})^2 \!+ \!(m_y \lambda/L_{s,y})^2 \le 1\}\!\!
\\\label{epsilon_r}
\hspace{-0.4cm}\mathcal{E}_r &= \{(\ell_x,\ell_y)\!\in\!\mathbb{Z}^2 \!:\! (\ell_x \lambda/{L_{r,x}})^2 + (\ell_y \lambda/L_{r,y})^2 \!\le 1\}\!
\end{align}
at the source and receiver, respectively.
We call $n_s = |\mathcal{E}_s|$ and $n_r=|\mathcal{E}_r|$ the cardinalities of the sets $\mathcal{E}_s$ and $\mathcal{E}_r$, respectively, and notice that they can be computed by counting the number of lattice points falling into the 2D lattice ellipses in~\eqref{epsilon_s} and~\eqref{epsilon_r}. For $\min(L_{s,x},L_{s,y})/\lambda \gg1$ and $\min(L_{r,x},L_{r,y})/\lambda \gg 1$, we have that~\cite{SPAWC}
\begin{align}\label{eq:ns_nr}
n_s =  \Big\lfloor{\frac{\pi}{\lambda^2}L_{s,x}L_{s,y}}\Big\rfloor \quad \quad
n_r = \Big\lfloor{\frac{\pi}{\lambda^2}L_{r,x}L_{r,y}}\Big\rfloor.
\end{align}
With the above definitions at hand, Theorem~\ref{th:series} is given.
\begin{theorem} \label{th:series}
The spatial impulse response $h(\vect{r},\vect{s})$ over rectangular volumetric arrays with finite side lengths $L_{r,x},L_{r,y}, L_{r,z} $ and $L_{s,x} ,L_{s,y},L_{s,z}$ can be approximated as $\min(L_{s,x},L_{s,y})/\lambda$ and $\min(L_{r,x},L_{r,y})/\lambda$ become large by the 4D Fourier plane-wave series in~\eqref{Fourier_series} where
\setcounter{equation}{24}
\begin{equation}
H_a(\ell_x,\ell_y,m_x,m_y)\sim\CN\Big(0,\sigma^2(\ell_x,\ell_y,m_x,m_y)\Big)
\end{equation}
 are independent, circularly-symmetric, complex-Gaussian random variables with variance given by 
\begin{equation}\label{variances}
 \frac{1}{(2\pi)^4}\!\! \iiiint_{\mathcal{S}_s(m_x,m_y) \times \mathcal{S}_{r}(\ell_x,\ell_y)} 
\hspace{-2.5cm}S(k_x,k_y,\kappa_x,\kappa_y)  dk_xdk_y d\kappa_xd\kappa_y
\end{equation}
where the sets $\mathcal{S}_s(m_x,m_y) $ and $\mathcal{S}_{r}(\ell_x,\ell_y)$ are defined as
\begin{align}\label{S_s}
  \!&\left\{\!\Big[\frac{2\pi m_x}{L_{s,x}},\frac{2\pi (m_x\!+\!1)}{L_{s,x}}\Big] \!\times \!\Big[\frac{2\pi m_y}{L_{s,y}},\frac{2\pi (m_y\!+\!1)}{L_{s,y}}\Big] \!\right\}\!\\\label{S_r}
 \!&\left\{\Big[\frac{2\pi \ell_x}{L_{r,x}},\frac{2\pi (\ell_x+1)}{L_{r,x}}\Big] \!\times\! \Big[\frac{2\pi \ell_y}{L_{r,y}},\frac{2\pi (\ell_y+1)}{L_{r,y}}\Big]\right\}
\end{align}
which are determined by the $xy-$spatial resolution of the volumes.
%
\end{theorem}
Theorem 2 shows that $h(\vect{r},\vect{s})$ can be approximately represented with an orthogonal discrete basis set of functions $\{\phi(\ell_x,\ell_y,m_x,m_y,\vect{r},\vect{s})\}$, which can be factorized as $\phi(\ell_x,\ell_y,m_x,m_y,\vect{r},\vect{s})= a_{r}(\ell_x,\ell_y, \vect{r})a_{s}(m_x,m_y,\vect{s})$.
%
These functions decompose the channel over the fixed angular sets $\mathcal{S}_s(m_x,m_y) \times \mathcal{S}_{r}(\ell_x,\ell_y)$ in the angular domain. Each variance $\sigma^2(\ell_x,\ell_y,m_x,m_y)$ in \eqref{variances} depends directly on the scattering as it represents the fraction of channel power transferred within every pair of angular sets.
%

%

We observe that the approximation error in~\eqref{Fourier_series}
reduces as $\min(L_{s,x},L_{s,y})/\lambda$ and $\min(L_{r,x},L_{r,y})/\lambda$ become large \cite{JSAC}, and vanishes
as $\min(L_{s,x},L_{s,y})/\lambda\to \infty$ and $\min(L_{r,x},L_{r,y})/\lambda\to \infty$ jointly. It thus follows that
\eqref{Fourier_series} does not require the volumes' size to be ``physically large'',
but rather their normalized size (with respect to $\lambda$). Remarkably, the approximation is accurate already for normalized sizes larger than a few wavelengths~\cite{JSAC}.
\section{Holographic MIMO Communications}
We now use the developed channel model in~\eqref{Fourier_series} to evaluate the capacity of a Holographic MIMO system with rectangular volume arrays of sub-wavelength spacing antenna elements. We denote by $N_s \ge n_s$ and $N_r\ge n_r$ the number of antennas at the transmitter and receiver, respectively. 
The $j$th transmit antenna is located at $\vect{s}_j$ whereas the $i$th receive antenna is located at $\vect{r}_i$. 
We call ${\bf u}_{s}(m_x,m_y)\in\mathbb{C}^{N_s}$ the transmit vector with entries $ \frac{1}{\sqrt{N_s}}a_{s}^*(m_x,m_y,\vect{s}_j)$ for $i=1,\ldots,N_s$ and denote ${\bf U}_s \in \mathbb{C}^{N_s \times n_s}$ the matrix collecting the $n_s$ column vectors $\{{\bf u}_{s}(m_x,m_y)\}$. Similarly, ${\bf u}_{r}(\ell_x,\ell_y)\in\mathbb{C}^{N_r}$ is the receive vector with entries $ \frac{1}{\sqrt{N_r}}a_{r}(\ell_x,\ell_y,\vect{r}_i)$  for $i=1,\ldots,N_r$ and ${\bf U}_r \in \mathbb{C}^{N_r\times n_r}$ is the matrix collecting the $n_r$ column vectors $\{{\bf u}_{r}(\ell_x,\ell_y)\}$. Since $\{a_{r}(\ell_x,\ell_y, \vect{r}_i)\}$ and $\{a_{s}(m_x,m_y,\vect{s}_j)\}$ constitute a set of orthonormal discrete basis functions on $\mathcal{V}_r$ and $\mathcal{V}_s$, respectively, we have that ${\bf U}_s$ and ${\bf U}_r$ are semi-unitary matrices for which ${\bf U}_s^{\Htran}{\bf U}_s = \vect{I}_{n_s}$ and ${\bf U}_r^{\Htran}{\bf U}_r = \vect{I}_{n_r}$. Their columns describe discretized transmit and receive plane-wave harmonics that span fixed angular sets.


From~\eqref{Fourier_series}, we can approximate the channel matrix $\vect{H} \in \mathbb{C}^{N_r\times N_s}$ obtained by spatial sampling $\mathcal{V}_r$ and $\mathcal{V}_s$ at points $\vect{r}_i$ and $\vect{s}_j$ for $i=1,\ldots,N_r$ and $j=1,\ldots,N_s$  as  
\begin{equation}\label{channel_matrix}
\vect{H} = {\bf U}_r
\vect{H}_a {\bf U}_s^{\Htran} \mathop{=}^{(a)}{\bf U}_r
\left(\boldsymbol{\Sigma} \odot \vect{W} \right){\bf U}_s^{\Htran}
\end{equation}
where $\vect{H}_a=\boldsymbol{\Sigma} \odot \vect{W} \in\mathbb{C}^{n_r\times n_s}$ collects the zero-mean and independent random entries $\{\sqrt{N_rN_s}H_a(\ell_x,\ell_y,m_x,m_y)\}$. Notice that $(a)$ follows by denoting $\vect{W}$ the matrix with i.i.d. circularly-symmetric Gaussian entries and $\boldsymbol{\Sigma}$ the matrix with entries $\{N_sN_r\sigma^2(\ell_x,\ell_y,m_x,m_y)\}$.
Because of its structure, \eqref{channel_matrix} constitutes a \emph{physics-based low-rank approximation} in UIU form~\cite{Tulino2005}, see also the model proposed by Weichselberger \emph{et al}. in~\cite{Weichselberger2006}.
In particular, due to the analogy between ${\bf U}_r$ and ${\bf U}_s$ and the discrete-Fourier transform matrix, the developed Fourier plane-wave framework can be considered as the extension to planar and volume arrays of the virtual channel model, pioneered in~\cite{Sayeed2002} for uniform linear arrays only.

%

The channel matrix~\eqref{channel_matrix} gives rise to correlated Rayleigh fading where the correlation matrix ${\bf R}_{\rm vec({\bf H})} = \mathbb{E}\{{\rm vec({\bf H})}{\rm vec}({\bf H})^{\Htran}\}$ of $\vect{H} $ takes the form 
\begin{equation}\label{correlation_matrix}
{\bf R}_{\rm vec({\bf H})}= ({\bf U}_s^*\kron{\bf U}_r)\mathbb{E}\{{\rm vec}({\bf H}_a){\rm vec}({\bf H}_a)^{\Htran}\}({\bf U}_s^{\Ttran}\kron{\bf U}_r^{\Htran})\end{equation}
where $\mathbb{E}\{{\rm vec}({\bf H}_a){\rm vec}({\bf H}_a)^{\Htran}\}$ is a diagonal matrix whose entries are $\{N_sN_r\sigma^2(\ell_x,\ell_y,m_x,m_y)\}$. These scaled variances are the eigenvalues of ${\bf R}_{\rm vec({\bf H})}$ and determine the correlations of ${\bf H}$. 
The more uneven these variances, the stronger the correlation. Ideally, if $\boldsymbol{\Sigma} = N_sN_r \vect{I}_{n_s n_r}$ the channel samples would be mutually independent, thus leading to the i.i.d. Rayleigh fading model.
However, this is never the case, except for half-wavelength spaced 1D arrays in isotropic scattering.
In fact, the variances $\{\sigma^2(\ell_x,\ell_y,m_x,m_y)\}$ are not all equal even in an isotropic scenario \cite{JSAC}, as it follows from using \eqref{psd} into \eqref{variances} by setting $ A(k_x,k_y,\kappa_x,\kappa_y) = \tilde A = \frac{2}{\kappa \eta} (\frac{2 \pi}{\sqrt{\kappa}})^2$. This proves that a stochastic channel model necessarily exhibits correlation~\cite{JSAC}, and makes it evident that the i.i.d. Rayleigh fading model shall never be used to model ${\bf H}$~\cite{bjornson2020rayleigh}. 
\subsection{Capacity evaluation}
Our MIMO system model is $\vect{y}  =  \sqrt{{\rm{snr}}}\vect{H} \vect{x} + \vect{z} $
where $\vect{y}$ and $\vect{x} $ are the input and output vectors, $\vect{z} \sim\CN({\bf 0},{\bf I}_{N_r})$ is white Gaussian noise and $\rm{snr}$ is the signal-to-noise ratio (SNR). Under the assumption of independent Gaussian inputs $\vect{x}  \sim\CN({\bf 0},\frac{1}{N_s}{\bf I}_{N_s})$, from the unitary invariance in~\eqref{channel_matrix} the ergodic channel capacity in bit/s/Hz is given by~\cite{Telatar}
\begin{align} \label{capacity_1}
\hspace{-0.3cm}C  
&=  \mathbb{E}\left\{\log_2 \det \left({\bf I}_{n_r} + \frac{{\rm snr} }{N_s}\vect{H}\vect{H}^{\Htran}\right)\right\}
\end{align}
from which, by using~\eqref{channel_matrix}, we obtain
%
\begin{align}
C =\sum_{i=1}^{{\rm rank}(\vect{H}_a)}\mathbb{E}\left\{\log_2 \left(1 + \frac{{\rm snr} }{N_s} \lambda_i\left(\vect{H}_a\vect{H}_a^{\Htran}\right)\right) \right\} \label{capacity_3}
\end{align}
where $\lambda_i\left(\vect{H}_a\vect{H}_a^{\Htran}\right)$ are the unordered positive eigenvalues of the matrix $\vect{H}_a\vect{H}_a^{\Htran}$. Since the entries of $\vect{H}_a$ are independent, with probability $1$ the rows and columns of $\vect{H}_a$ are linearly independent except for those that are identically zero. We denote $n_r^\prime \le n_r$ and $n_s^\prime \le n_s$ the number of rows and columns of $\vect{H}_a$ that are not identically zero. Hence, ${\rm rank}(\vect{H}_a) = \min(n_s',n_r')$ and the degrees of freedom (DoF) in~\eqref{capacity_3} are
\begin{equation} \label{DoF}
n_{\rm DoF} = \min(n_s',n_r') \le \min(n_s,n_r)
\end{equation}
where $n_s,n_r$ are given in~\eqref{eq:ns_nr}. Since $H_a(\ell_x,\ell_y,m_x,m_y)\sim\CN(0,\sigma^2(\ell_x,\ell_y,m_x,m_y))$, $n_{\rm DoF}$ equals the upper bound when non-zero power is received from all angular sets. 
In this case, the DoF per m$^2$ are given by $\pi/\lambda^2$ as it follows from~\eqref{eq:ns_nr}. If not, the DoF may eventually reduce.
%
%
%


\subsection{Numerical analysis}

Numerical simulations are now used to evaluate the channel capacity with the developed channel model. We assume that that the transmitter and receiver make use of two identical squared rectangular arrays with equal side length $L$. 
We assume that ${L=1}$~m and ${\lambda= 0.1}$~cm (i.e., ${L/\lambda=10}$). From~\eqref{eq:ns_nr}, we have that ${n_s=n_r=314}$. For simplicity, we consider a scattering propagation scenario with unit average power that is separable and symmetric between transmitter and receiver, i.e., $A_s(\kappa_x,\kappa_y) = A_r(k_x,k_y)$ and $\sigma_r^2(\ell_x,\ell_y)=\sigma_s^2(m_x,m_y)$ with
$\sigma^2(\ell_x,\ell_y,m_x,m_y) = \sigma_r^2(\ell_x,\ell_y)\sigma_s^2(m_x,m_y)$. At the receiver, we have
\begin{align}  \label{variances_rx}
  \sigma_r^2(\ell_x,\ell_y) &= \frac{1}{(2\pi)^2}\!\! \iint_{\mathcal{S}_r(\ell_x,\ell_y) } \hspace{-0.75cm}{\kappa \eta}/{2}\frac{A_r(k_x,k_y) }{\gamma(k_x,k_y)}
  d\kappa_xd\kappa_y
\end{align}
as it follows from using \eqref{psd} into \eqref{variances} under separability. 
To model the spectral factor $A_r(\kappa_x,\kappa_y)$, we use a 3D Von Mises-Fisher angular distribution function 
that in spherical coordinates ${(\kappa_x,\kappa_y)\!\to\!(\theta,\phi)}$ takes the form~\cite{Mammasis} 
\begin{equation} \label{VMF_pdf}
\frac{\alpha e^{(\alpha \sin(\theta)\sin(\mu_\theta) \cos(\phi-\mu_\phi) + \cos(\theta) \cos(\mu_\theta))} \sin(\theta)}{4\pi \sinh(\alpha)}
\end{equation}
where $\mu_\theta,\mu_\phi$ represent the mean directions and $\alpha$ is the so-called concentration parameter. We fix $\mu_\theta =\mu_\phi= 30^\circ$ and set $\alpha$ such that the circular variance is $\nu = 30^\circ$ in azimuth and elevation. 
\begin{figure} [t!]
        \centering
        \begin{subfigure}[t]{\columnwidth} \centering 
	\begin{overpic}[width=0.85\columnwidth,tics=10]{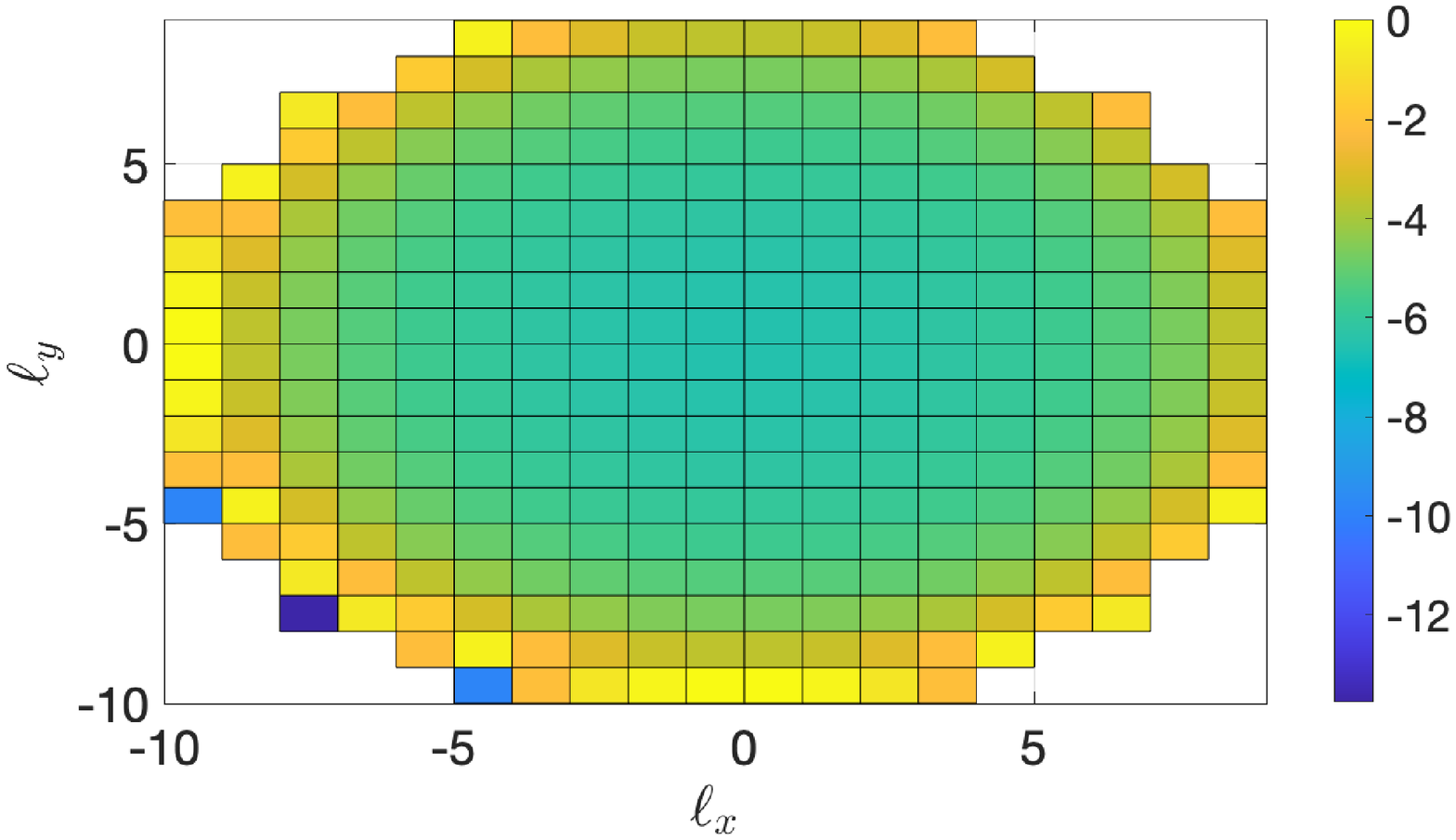}
\end{overpic} \vspace{-0.2cm}
                \caption{Isotropic propagation $\nu = 360^\circ$} \vspace{0.03cm}
                \label{fig:varIso} 
        \end{subfigure} 
        \begin{subfigure}[t]{\columnwidth} \centering  
	\begin{overpic}[width=0.85\columnwidth,tics=10]{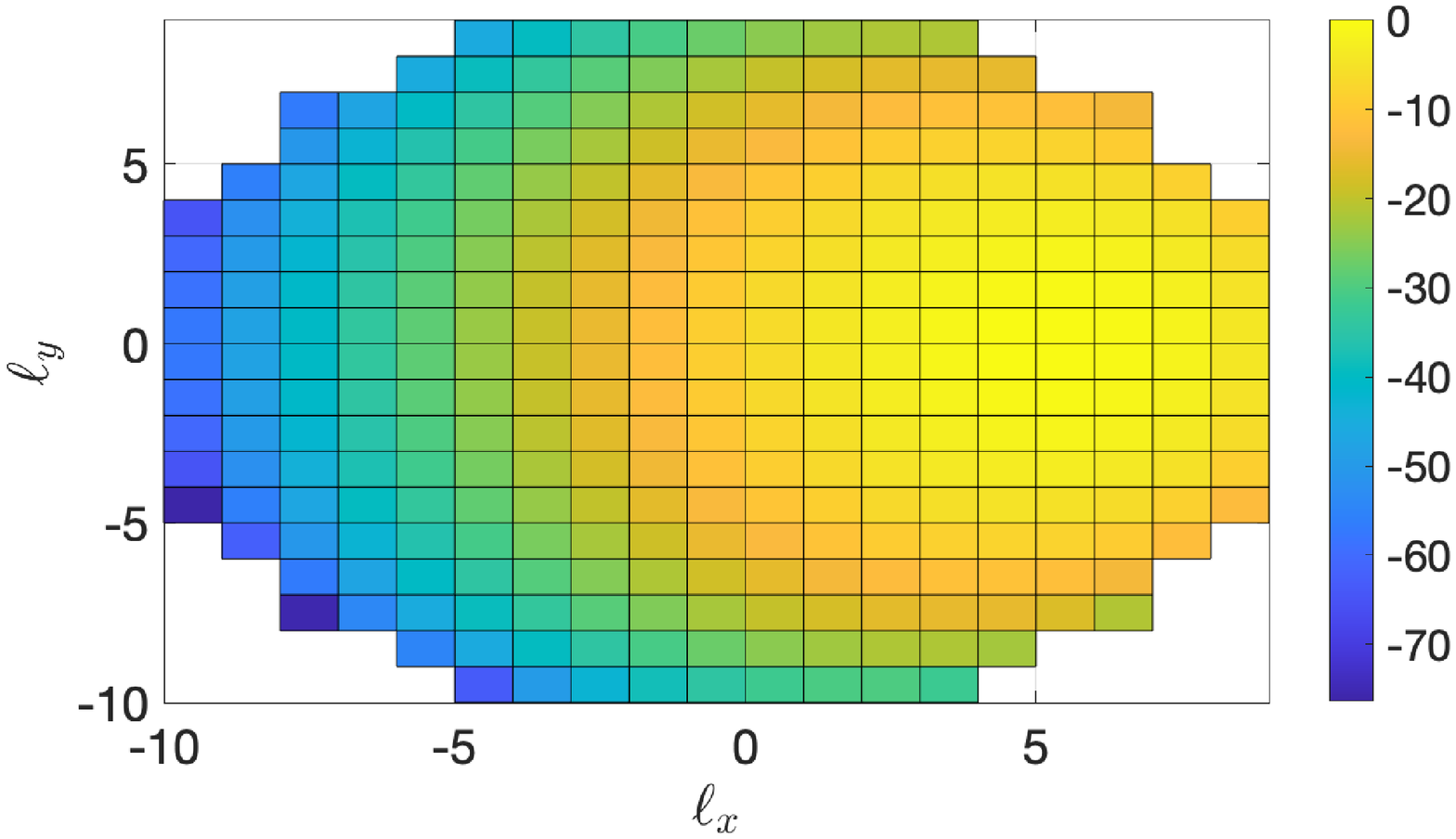}
\end{overpic}  \vspace{-0.2cm}
                \caption{Non-isotropic propagation $\nu = 30^\circ$} 
                \label{fig:varNonIso}
        \end{subfigure}\vspace{0.1cm}
        \caption{Normalized variances $\sigma_r^2(\ell_x,\ell_y)$ in dB for isotropic and non-isotropic propagations. The non-isotropic variances are generated by using the angular distribution function \eqref{VMF_pdf}.}
        \label{fig:eigenvalues}\vspace{-0.6cm}
\end{figure}

Fig.~\ref{fig:eigenvalues} shows the normalized variances $\sigma_r^2(\ell_x,\ell_y)$ for the ${n_r = 314}$ angular sets under isotropic and non-isotropic propagation conditions. In the latter case, they are computed by plugging \eqref{VMF_pdf} into \eqref{variances_rx}. {In both scenarios, the variances are non-zero only within the lattice ellipse $\mathcal{E}_r$ in \eqref{epsilon_r}.} Fig.~\ref{fig:varIso} confirms that $\{\sigma_r^2(\ell_x,\ell_y)\}$ are not all exactly equal under isotropic conditions (i.e., the channel exhibits spatial correlation). Particularly, they have a bowl-shaped behavior, notice that they are always bounded since the integrand in \eqref{variances_rx} is singularly-integrable \cite{JSAC}. For the propagation scenario in Fig.~\ref{fig:varNonIso}, the variances achieves higher values around the angular directions specified by \eqref{VMF_pdf}. 

\begin{figure}[t!]
    \centering
     \includegraphics[width=.95\columnwidth]{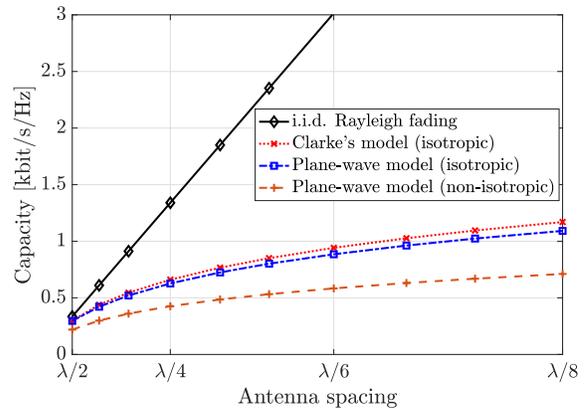} \vspace{-0.2cm}
  \caption{Ergodic capacity $C$ as a function of antenna spacing for ${\rm snr} = 0$~dB. The non-isotropic channel model is generated by using the angular distribution function in \eqref{VMF_pdf}.}
   \label{fig:capacity}\vspace{-0.5cm}
\end{figure}

Fig.~\ref{fig:capacity} plots the ergodic capacity as a function of the antenna spacing $\Delta\in[\lambda/2,\lambda/8]$ under both conditions of Fig.~\ref{fig:eigenvalues}. The ergodic capacity is evaluated by using the large-dimensional approximation in~\cite{Tulino2005}. Comparisons are made with the 3D Clarke's isotropic model. The i.i.d. Rayleigh fading model is used as reference. Fig.~\ref{fig:capacity} shows that the capacity of the plane-wave model in \eqref{Fourier_series} is very close to that of the Clarke's model for the setting of Fig.~\ref{fig:eigenvalues}a. {This validates our physical low-rank approximation in \eqref{channel_matrix} under isotropic propagation conditions.} Compared to i.i.d. fading, a large gap is observed for $\Delta \le \lambda/2$. This is due to the correlation that naturally arises among antennas when $\Delta$ decreases. The larger the correlation, the larger the gap. This is made evident by the capacity achieved under the non-isotropic conditions of Fig.~\ref{fig:eigenvalues}b.

%
%

\section{Summary}
We derived a 4D plane-wave series representation in Cartesian coordinates of a spatially-stationary scattering channel. This representation captures the essence of electromagnetic propagation and provides an angular-domain decomposition of the channel, which is valid for rectangular volumetric arrays. The developed model was used to evaluate the capacity of Holographic MIMO systems and to show that the i.i.d. Rayleigh fading model shall never be used for such systems.

%
%


\bibliographystyle{IEEEtran}
\bibliography{IEEEabrv,refs}

\end{document}